\documentclass[twocolumn]{aastex631}

\usepackage{multirow}

\usepackage{epsfig}
\usepackage{epsfig}  
\usepackage{amsmath} 
\usepackage{amssymb}
\usepackage{multirow}
\usepackage{natbib} 
\usepackage{hyperref}
\usepackage{graphicx} 
\usepackage{endnotes}
\usepackage{color}

\usepackage{graphicx}
\usepackage{amsmath}	
\usepackage{amssymb}	
\usepackage{ulem}
\usepackage{tabularx,booktabs}

\definecolor{Lightblue}{rgb}{0.68,0.85,0.9}

\definecolor{magenta}{rgb}{0.8, 0.6, 0.8}

\definecolor{orange}{rgb}{0.93, 0.57, 0.13}

\shorttitle{Polarization in galactic ULX Pulsar Swift J$0243.6+6124$}
\shortauthors{Majumder et al.}

\graphicspath{{./}{figures/}}

\begin{document}

\title{First detection of X-ray polarization in Galactic ULX pulsar Swift J0243.6$+$6124 with {\it IXPE}}

\author[0000-0002-2214-35937]{Seshadri Majumder}
\affiliation{Indian Institute of Technology Guwahati, Guwahati, 781039, India.}

\author{Rwitika Chatterjee}
\affiliation{Space Astronomy Group, ISITE Campus, U. R. Rao Satellite Centre, Outer Ring Road, Marathahalli, Bangalore, 560037, India.}

\author{Kiran M. Jayasurya}
\affiliation{Space Astronomy Group, ISITE Campus, U. R. Rao Satellite Centre, Outer Ring Road, Marathahalli, Bangalore, 560037, India.}

\author[0000-0003-4399-5047]{Santabrata Das}
\affiliation{Indian Institute of Technology Guwahati, Guwahati, 781039, India.}

\author{Anuj Nandi}
\affiliation{Space Astronomy Group, ISITE Campus, U. R. Rao Satellite Centre, Outer Ring Road, Marathahalli, Bangalore, 560037, India.}

\begin{abstract}
		We report the results of first ever spectro-polarimetric analyses of the Galactic ultra-luminous X-ray pulsar Swift J0243.6$+6124$ during the 2023 outburst using quasi-simultaneous {\it IXPE}, {\it NICER} and {\it NuSTAR} observations. A pulsation of period $\sim 9.79$ s is detected in {\it IXPE} and {\it NuSTAR} observations with pulse fractions (PFs) $\sim 18\%$ ($2-8$ keV) and $\sim 28\%$ ($3-78$ keV), respectively. Energy dependent study of the pulse profiles with {\it NuSTAR} indicates an increase in PF from $\sim 27\%$ ($3-10$ keV) to $\sim 50\%$ ($40-78$ keV). Further, epoch-dependent polarimetric measurements during the decay phase of the outburst confirm the detection of significant polarization, with the polarization degree (PD) and polarization angle (PA) ranging between $\sim 2 -3.1\%$ and $\sim8.6^{\circ}-10.8^{\circ}$, respectively, in the $2-8$ keV energy range. We also observe that the PD increases up to $\sim 4.8\%$ at higher energies ($\gtrsim 5$ keV) with dominating \texttt{bbodyrad} flux contribution ($1.5 \lesssim F_{\rm BB}/F_{\rm PL} \lesssim 3.4$) in the {\it IXPE} spectra. The phase-resolved polarimetric study yields PD as $\sim 1.7-3.1\%$ suggesting a marginal correlation with the pulse profiles. Moreover, the broad-band ($0.6-70$ keV) energy spectrum of combined {\it NICER} and {\it NuSTAR} observations is well described by the combination of \texttt{bbodyrad} and \texttt{cutoffpl} components with seed photon temperature ($kT_{\rm bb}$) $\sim 0.86 \pm 0.03$ keV and photon index ($\Gamma$) $\sim 0.98 \pm 0.01$. With the above findings, we infer that the observed `low' PD in Swift J0243.6$+6124$ is attributed possibly due to `vacuum resonance' effect between the overheated and relatively cooler regions of the neutron star boundary layer.

\end{abstract}


\section{Introduction} \label{sec:intro}

Ultra-luminous X-ray pulsar sources (ULXPs) are bright point-like off-nuclear objects containing neutron stars with isotropic luminosity ($\sim 10^{39-41}$ erg $\rm s^{-1}$) exceeding the Eddington limit \cite[for review]{Feng-Soria2011,King-etal2023}. Alongside the black hole candidates \cite[]{Atapin-etal2019,Majumder-etal2023}, the detection of $1.4$ s X-ray pulsation in M82 X$-2$ \citep{Bachetti-etal2014} opened up a completely new window to understand the characteristics of ULXs. Meanwhile, nine confirmed ULXPs are reported with the detection of X-ray pulsation \cite[and references therein]{King-etal2023} till date.

Admittedly, the accretion onto the X-ray pulsars (XRPs) is regulated by strong magnetic fields ($\sim 10^{12-13}$ G). The accreted matter is channelled along the magnetic field lines at the magnetospheric radius and generates hotspots near the magnetic poles of the neutron star (NS) that radiates pulsed emission in X-rays being misaligned with the spin axis (see \citealt[for a recent review]{Mushtukov-etal2022}). It is believed that at higher mass accretion rates, the hotspots turn into vertically extended accretion columns above the NS surface \citep{Basko-etal1976, Mushtukov-etal2015}. However, relative contribution in total emission from the accretion column, depending on the disc truncation radius, is found to play a vital role for the irregular pulsations detected in most of the known ULXPs \citep{Walton-etal2018}. Indeed, ULXP spectra are empirically well described by blackbody-like emissions and power-law profiles with high energy cut-off.

\begin{figure*}
	\begin{center}
		\includegraphics[width=\textwidth]{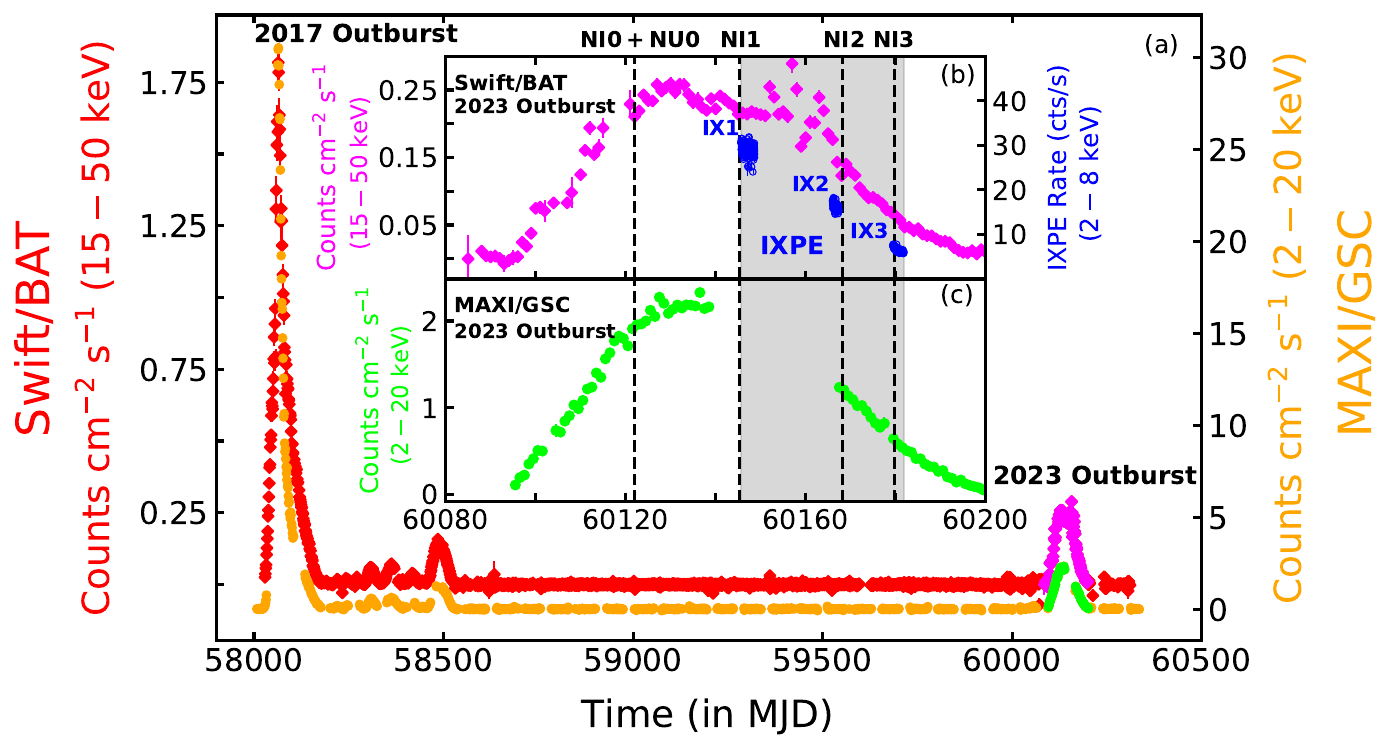}
	\end{center}
	\caption{Panel (a): 1 day binned {\it MAXI/GSC} (orange) and {\it Swift/BAT} (red) light curves of Swift J0243.6$+6124$ since its discovery (October, 2017) in $2-20$ keV and $15-50$ keV energy bands. Panel (b) and (c): Zoomed view of the 2023 outburst observed with {\it Swift/BAT} (magenta) and {\it MAXI/GSC} (green). Epochs of {\it NICER} and {\it NuSTAR} observations are marked with vertical lines. Grey patch indicates the duration of {\it IXPE} observations and blue circles denote the corresponding {\it IXPE} count rates of 1000 s bin combining all three DUs ($2-8$ keV). 
		}
	\label{fig:swift_BAT}
\end{figure*}

The {\it Imaging X-ray Polarimetry Explorer} \cite[{\it IXPE};][]{Weisskopf-etal2022} provides an unique opportunity to probe the X-ray polarization of XRPs. So far, the detection of phase-averaged/resolved polarized emission in a handful of XRPs, such as Her X$-1$ ($\sim 10\%$, \citealt{Doroshenko-etal2022}), Cen X$-3$ ($\sim 5.8\%$, \citealt{Tsygankov-etal2022}), 4U $1626-67$ ($\sim 4.8\%$, \citealt{Marshall-etal2022}), Vela X$-1$ ($\sim 2.3\%$, \citealt{Forsblom-etal2023}),  GRO J$1008-57$ ($\sim 15\%$, \citealt{Tsygankov-etal2023}), EXO $2030+375$ ($\sim 3\%$, \citealt{Malacaria-etal2023}), X Persei ($\sim 20\%$, \citealt{Mushtukov-etal2023}) and GX $301-2$ ($\sim 3-10\%$, \citealt{Suleimanov-etal2023}) is confirmed with {\it IXPE}. If not all, most of the sources show `low' polarization degree (PD) compared to the predictions from the existing models \citep{Meszaros-etal1988,Caiazzo-etal2021b,Caiazzo-etal2021a}, and its cause remains an open question till date. 

\begin{figure*}
	\begin{center}
		\includegraphics[width=0.65\columnwidth]{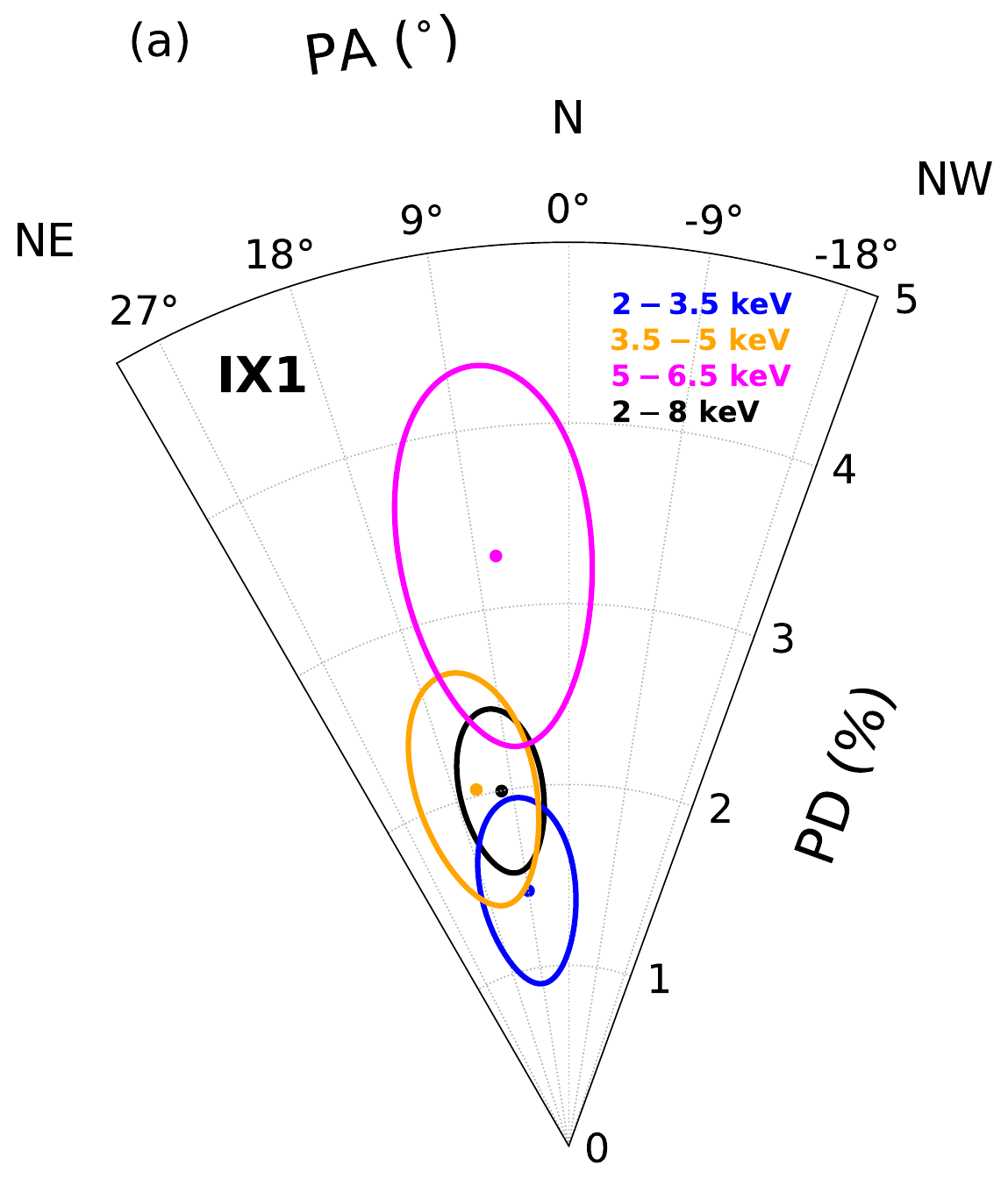}
		\includegraphics[width=0.65\columnwidth]{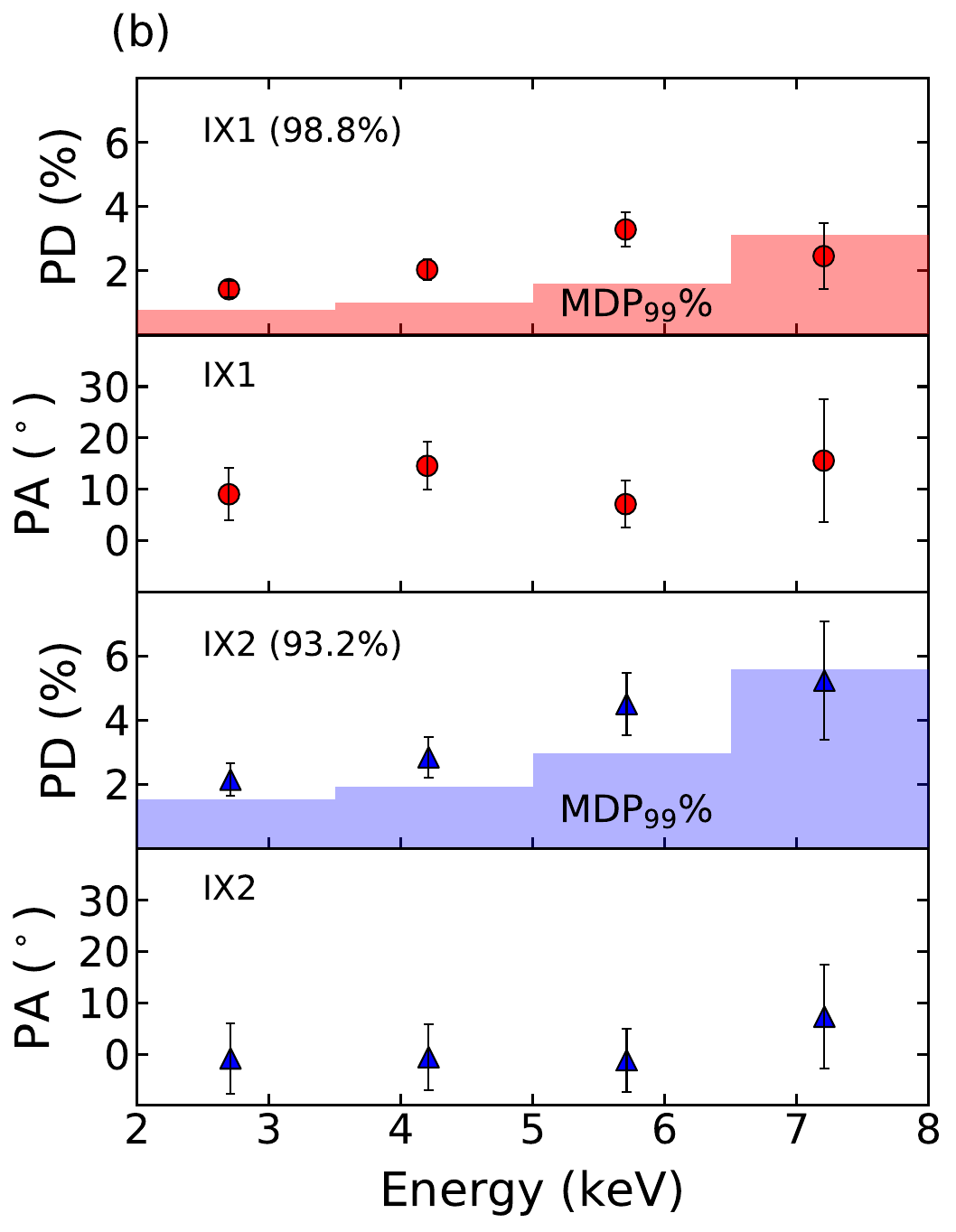}
		\includegraphics[width=0.7\columnwidth]{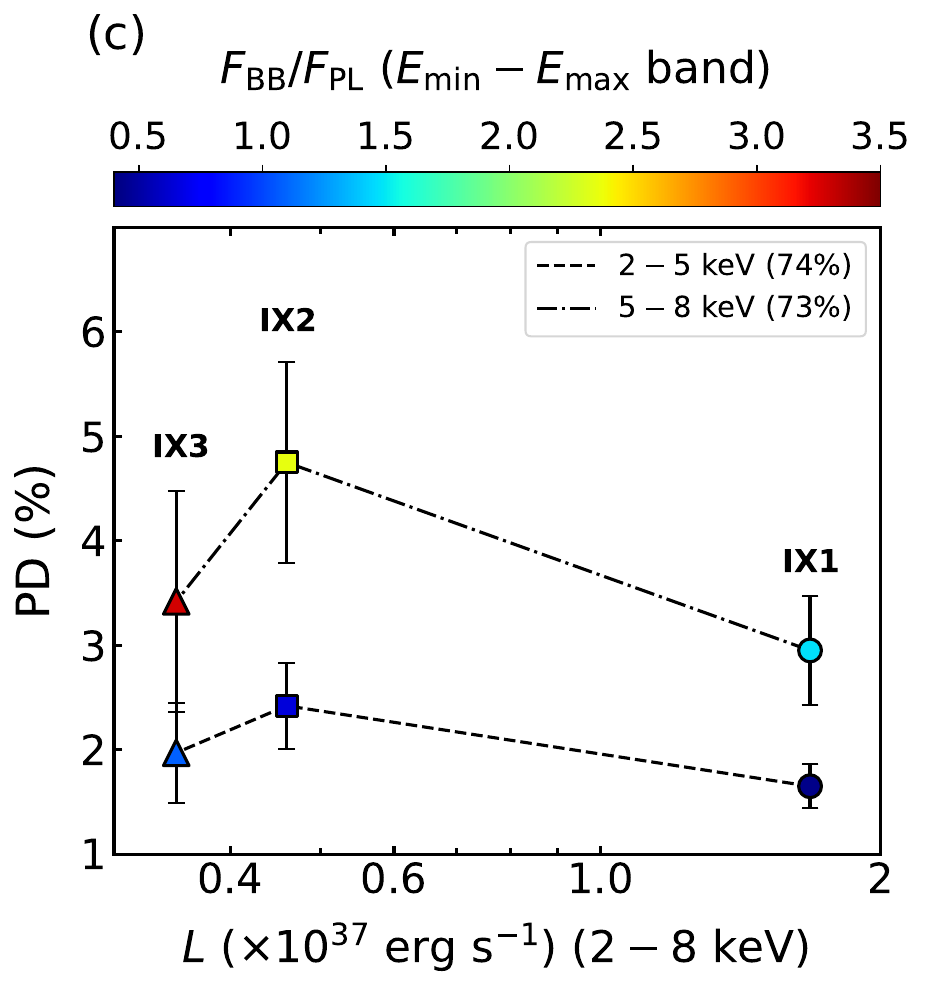}
	\end{center}
	\caption{{\it Left}: Confidence contours ($2\sigma$) of PD and PA obtained in epoch IX1 are shown with different colors for different energy bands. {\it Middle}: Variations of PD and PA with energy for epoch IX1 (MJD$_{\rm start}$ $60145.63$) and IX2 (MJD$_{\rm start}$ $60165.99$). Histograms denote $\rm MDP_{99}\%$, and confidence levels ($\%$) for PD variations with energy are mentioned in the legend. {\it Right}: Correlation of PD (model-independent) with luminosity and flux ratio ($F_{\rm BB}/F_{\rm PL}$) from spectro-polarimetric modelling of {\it IXPE} data. PD variations are significant at $74\%$ and $73\%$ in the respective energy bands.}
	\label{fig:ixpe_pol}
\end{figure*}

Indeed, most of the aforementioned models were developed neglecting the possible effects of specific temperature profiles at the NS surface. It is important to note that due to accretion, the overheated upper boundary layer of the NS surface can significantly alter the properties of polarized emission \citep{Tsygankov-etal2022}. In addition, the complex magnetic field geometry perhaps causes the mixing of emissions from different parts of the NS surface which could possibly result in `low' PD in EXO $2030+375$ \citep{Malacaria-etal2023}. Furthermore, the scattering and reprocessing of X-ray emissions in the surrounding stellar wind of the companion can marginally depolarize the intrinsic emission up to $10-15\%$ \citep{Suleimanov-etal2023}. Notably, significant phase dependent polarization properties are often observed in XRPs despite the low phase-averaged measurements. For example, GX $301-2$ exhibits a polarization of $\sim 3-10\%$ over different pulse phases, whereas the phase-averaged estimate results in a null detection \citep{Suleimanov-etal2023}. Similar findings are also observed in GRO J$1008-57$ \citep{Tsygankov-etal2023} and Vela X$-1$ \citep{Forsblom-etal2023}. 

In this work, we study the polarization properties of the transient XRP Swift J$0243.6+6124$ for the first time, using {\it IXPE} observations of this source. The source, having a Be-star binary companion \cite[]{Reig-etal2020}, was discovered \cite[]{Kennea-etal2017} during its giant outburst in 2017 by {\it Swift/BAT} with a peak flux of $\sim 8.2$ Crab and pulse period of $\sim 9.86$ s \cite[]{Kennea-etal2017}. With the measured source distance of $6.8$ kpc \cite[]{Bailer-Jones-etal2018}, the peak luminosity of the source Swift J0243.6 + 6124, classified as the first Galactic ULXP \cite[]{Tsygankov-etal2018}, exceeds the Eddington limit of a NS system.

In this Letter, we present the results of in-depth phase-averaged/resolved spectro-polarimetric analyses of Swift J0243.6$+6124$ with {\it IXPE} in $2-8$ keV energy range. In addition, we also use quasi-simultaneous {\it NICER} and {\it NuSTAR} observations to study the broad-band ($0.6-70$ keV) spectral distribution of the source.

The Letter is organized as follows: In \S 2, we mention the observation details along with the data reduction procedures of each instrument. In \S 3, we present the results obtained from the spectro-polarimetric studies. Finally, we summarize our findings and conclude in \S 4.

\section{Observation and Data Reduction} \label{s:Data-reduction}

\begin{figure*}
    \begin{center}
    \includegraphics[scale=0.52]{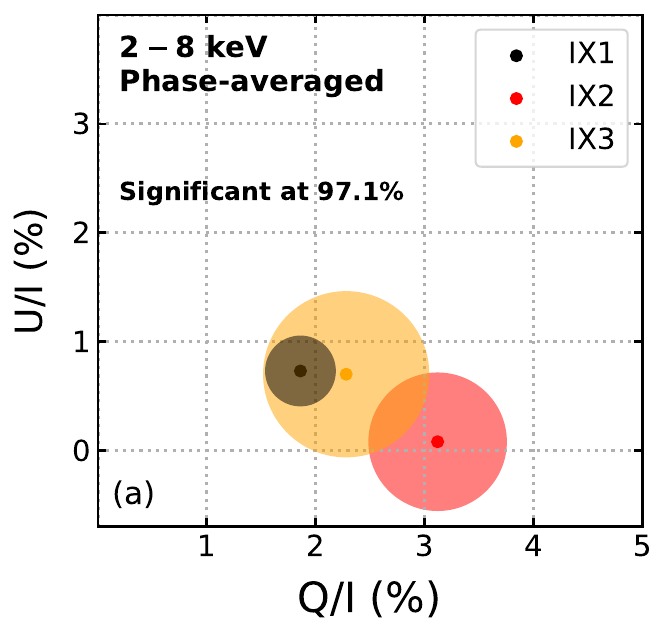}
    \includegraphics[scale=0.52]{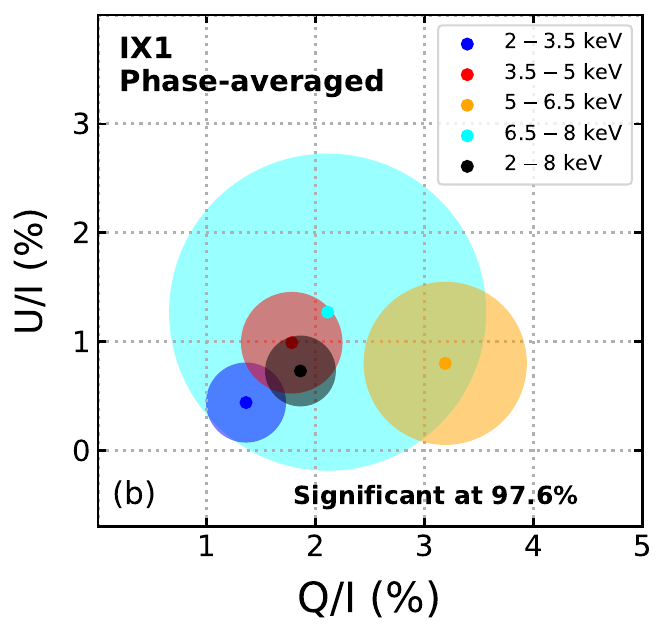}
    \includegraphics[scale=0.52]{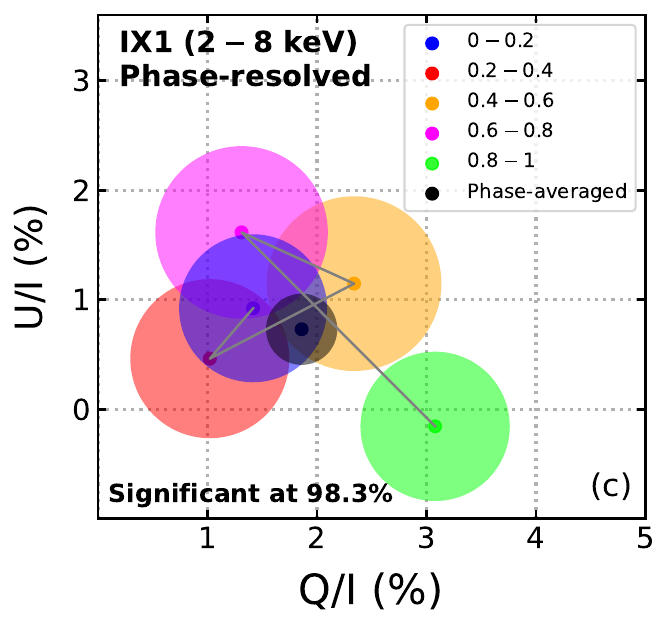}
    \end{center}
	\caption{Normalized stokes parameters (Q/I and U/I) obtained with \texttt{PCUBE} analyses for (a) different epochs ($2-8$ keV), (b) energy bands (epoch IX1) and (c) phase bins (epoch IX1 in $2-8$ keV). The radii of colored circles represent $1 \sigma$ uncertainty values corresponding to two degrees of freedom. See the text for details.
	}
	\label{fig:stokes}
\end{figure*}

{\it IXPE} observed Swift J0243.6$+6124$ three times between July 20, 2023 to August 25, 2023 for a total exposure of about $\sim 375$ ks during the decay phase of its 2023 outburst. The entire observation is segmented into three epochs of $\sim 167$ ks (IX1, $\rm MJD_{\rm start}$ 60145.63), $\sim 77$ ks (IX2, $\rm MJD_{\rm start}$ 60165.99) and $\sim 131$ ks (IX3, $\rm MJD_{\rm start}$ 60179.42) exposures, respectively. We make use of cleaned and calibrated level-2 event files from the three detector units (DUs) of {\it IXPE} ($2-8$ keV). The data analysis is carried out using \texttt{IXPEOBSSIMv30.5.0} software \citep{Baldini-etal2022} following standard procedures mentioned in \citealt{Kislat-etal2015, Strohmayer-etal2017, Kushwaha-etal2023, Majumder-etal2024}. The source and background regions are considered as the $60^{\prime\prime}$ circular region at the source coordinate and the annular region between $180^{\prime\prime}$ and $240^{\prime\prime}$ radii with the same center, respectively \cite[see also][]{Jayasurya-etal2023, Majumder-etal2024}. Further, \texttt{XPSELECT} task is used to extract the source and background events from the selected regions. We use \texttt{XPBIN} task with various algorithms such as \texttt{PCUBE}, \texttt{PHA1}, \texttt{PHA1Q} and \texttt{PHA1U} to generate necessary data products for model-independent \citep{Kislat-etal2015} and spectro-polarimetric \citep{Strohmayer-etal2017} studies. Finally, \texttt{XPPHASE} task is used to assign phase to barycenter-corrected\footnote{\url{https://heasarc.gsfc.nasa.gov/ftools/caldb/help/barycorr.html}} IXPE events lists for phase-resolved polarimetric studies.

Swift J0243.6$+6124$ is also observed by {\it NICER} and {\it NuSTAR} during the 2023 outburst. In this work, we analyze quasi-simultaneous {\it NICER} ($\sim 4$ ks) and {\it NuSTAR} ($\sim 12$ ks) observations (hereafter NI0 and NU0) carried out on June, 27, 2023. Additionally, we consider multiple {\it NICER} observations (hereafter NI1, NI2, NI3) which are quasi-simultaneous with the three {\it IXPE} epochs (IX1, IX2, IX3), respectively. The details of all the multi-mission observations along with their exposures, used in this work, are tabulated in Table \ref{table:obs_log}. The data is processed using standard analysis softwares \texttt{NICERDASv11a} and \texttt{nupipelinev0.4.9} for {\it NICER} and {\it NuSTAR}, respectively, integrated in \texttt{HEASOFT V6.32.1}\footnote{\url{https://heasarc.gsfc.nasa.gov/docs/software/heasoft}}. We use appropriate calibration databases while analysing data.

\section{Analysis and Results} \label{s:results}

\subsection{Outburst Profile} \label{s:outburst}

Swift J0243.6$+6124$ has been monitored by {\it MAXI/GSC} ($2-20$ keV) and {\it Swift/BAT} ($15-50$ keV) almost on a daily basis since the trigger of its giant outburst in $2017$. We present the complete coverage of the source with different instruments since its detection in panel (a) of Fig. \ref{fig:swift_BAT}. The recent outburst in $2023$ covered by both {\it Swift/BAT} and {\it MAXI/GSC} is shown in panels (b) and (c) of Fig. \ref{fig:swift_BAT}, respectively. We observe almost similar profile of the rise and decay phases of the outburst with peak {\it MAXI/GSC} and {\it Swift/BAT} flux of about $2.27$ counts $\rm cm^{-2}$ $\rm s^{-1}$ ($597$ mCrab) and $\sim 0.26$ counts $\rm cm^{-2}$ $\rm s^{-1}$ ($1.2$ Crab) in $2-20$ keV and $15-50$ keV energy ranges, respectively. The background-subtracted {\it IXPE} light curves of 1000 s bin combining all three DUs ($2-8$ keV) during epochs IX1 ($\sim 973$ mCrab), IX2 ($\sim 843$ mCrab) and IX3 ($\sim 314$ mCrab) are shown in panel (b) using blue open circles along with the entire observation period in grey shade. The quasi-simultaneous {\it NICER} and {\it NuSTAR} observations over different epochs are marked with vertical dashed lines. We observe that the average {\it IXPE} count rate gradually decreases as $29.46$ cts/s (IX1), $16.47$ cts/s (IX2) and $6.55$ cts/s (IX3) during the decay phase of the outburst.

\subsection{Phase-averaged Polarimetric Measurements} \label{s:polar}

\subsubsection{Model-independent \texttt{PCUBE} Results}

\begin{figure*}
    \begin{center}
    \includegraphics[scale=0.5]{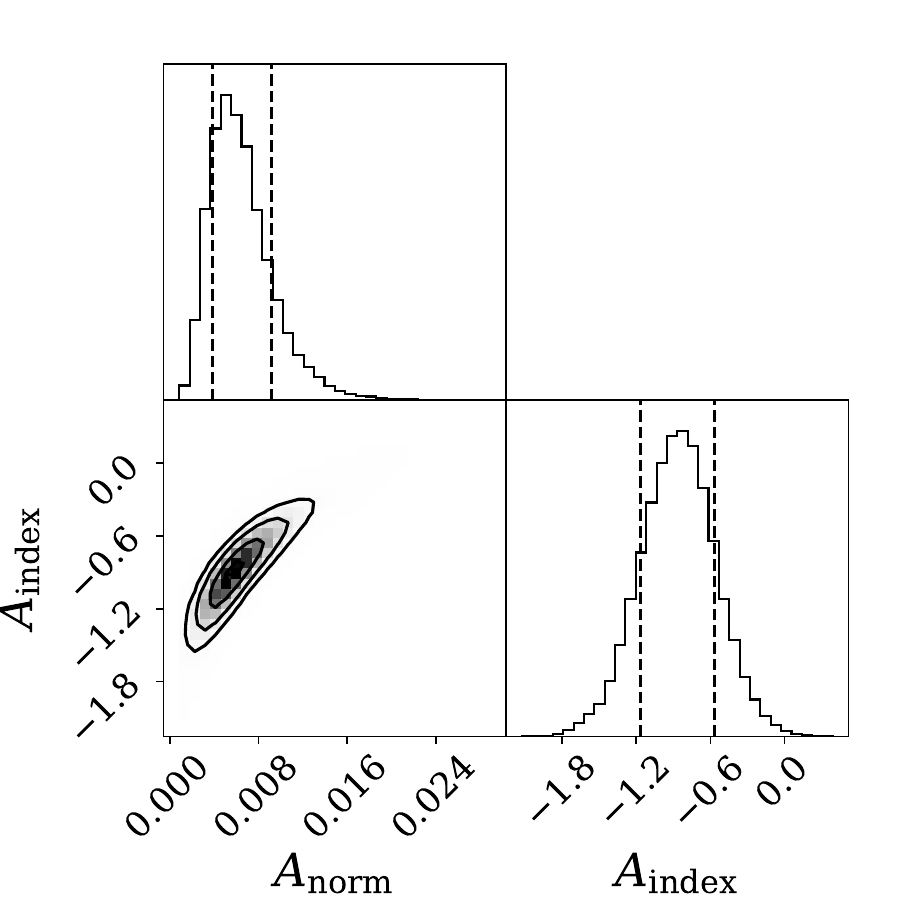}
    \includegraphics[scale=0.4]{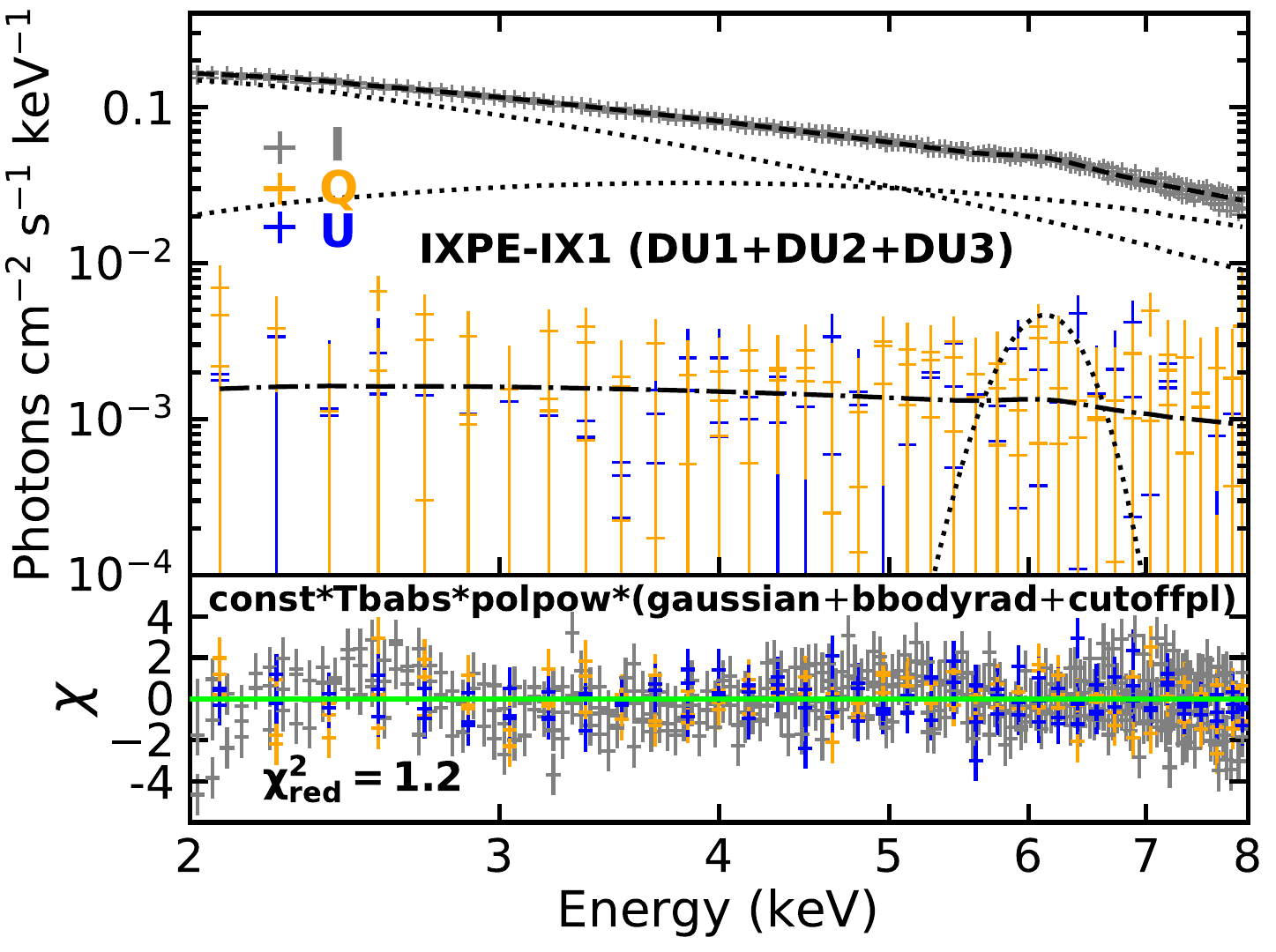}
    \end{center}
    \caption{{\it Left}: Corner plot shows the covariance between best-fitted $A_{\rm index}$ and $A_{\rm norm}$ parameters of \texttt{polpow} model in $2-8$ keV energy band, obtained using MCMC method. The contours represent $1\sigma$, $2\sigma$ and $3\sigma$ confidence range, respectively. {\it Right}: Best-fitted I, Q and U Stokes spectra of {\it IXPE} combining all DUs in $2-8$ keV energy range ({\it top panel}) and corresponding residual variation ({\it bottom panel}). See the text for details.}
    \label{fig:spec-pol}
\end{figure*}

For model-independent analysis, we use the \texttt{PCUBE} algorithm to estimate normalized Stokes parameters, polarization angle (PA), polarization degree (PD) and minimum detectable polarization at $99\%$ confidence ($\rm MDP_{99}\%$) following \citealt[and references therein]{Majumder-etal2024}. Considering all events from three detector units (DUs) of {\it IXPE}, we measure the polarization parameters during epoch IX1 as PD $=2.0\pm0.2\%$ ($> 8 \sigma$), PA $=10.8^\circ\pm3.3^\circ$ with $\rm MDP_{99} = 0.7\%$ in $2-8$ keV energy band. We also detect significant polarization ($\rm PD > \rm MDP_{99}$, $\sigma > 3$) during epochs IX2 and IX3, respectively. We observe that PD increases to $\sim 3.1 \pm 0.5\%$ ($> 6 \sigma$) in epoch IX2, which subsequently decreases to $2.4\pm0.5\%$ ($> 3 \sigma$) in epoch IX3. Note that the constrained PA remains $0.7^\circ\pm4.2^\circ$ in epoch IX2 and becomes $8.9^\circ\pm6.4^\circ$ during epoch IX3. Following \cite{Krawczynski-etal2022}, we estimate the significance of the PD variation in different epochs considering the $\chi^{2}$ statistics of 2 degrees of freedom. Here, we consider the null hypothesis as PD$=2.2\pm0.2\%$, obtained from the data combining all the epochs. With this, we find that the change in PD over the epochs is significant at $92\%$ confidence level, indicating a marginal variation. In Fig. \ref{fig:stokes}a, we present the corresponding variation of the normalized Stokes parameters (Q/I and U/I) over different epochs along with $1\sigma$ contours in Q-U space. Furthermore, we estimate the significance of the variations of normalized Stokes parameters (Q/I and U/I) over different epochs against the respective averaged out values, considering $2(n-1)$ degrees of freedom, $n$ being the number of Q/I$-$U/I pairs. We find that the variation in Q/I and U/I over three {\it IXPE} epochs is significant at $97.1\%$ confidence within $2-8$ keV energy range, which is higher than the obtained significance ($92\%$) of the corresponding variation in PD. This is expected because of the independent and Gaussian nature of the error distributions associated with the Stokes parameters.

Further, to infer the energy dependent polarimetric properties, we estimate polarization parameters in different energy bands, namely $2-3.5$ keV, $3.5-5$ keV, $5-6.5$ keV and $6.5-8$ keV, respectively. We observe an increase of PD from $\sim 1.4\%$ ($2-3.5$ keV) to a maximum of $\sim 3.3\%$ ($5-6.5$ keV) with a null-detection in $6.5-8$ keV energy band during epoch IX1. We estimate the significance of the energy dependent PD variation in each epoch following the approach discussed above. In doing so, we consider the $\chi^2$ statistics of $3$ degrees of freedom (corresponding to four energy bins) and compute the significance against the PD in $2-8$ keV energy band (see Table 2) of respective epochs. We find that the increase of PD with energy is significant at $98.8\%$ (IX1), $93.2\%$ (IX2) and $75.5\%$ (IX3) confidence level for the individual epochs. Based on these results, we indicate that the PD moderately increases with energy for IX1, whereas marginal variation with energy is observed for IX2 and IX3, respectively. However, the energy dependence of PA is found to be insignificant. In Fig. \ref{fig:ixpe_pol}a, we present $2\sigma$ confidence contours of PD and PA in different energy ranges for epoch IX1. The obtained energy dependent behavior of PD and PA at different epochs (IX1 and IX2) is shown in Fig. \ref{fig:ixpe_pol}b. The estimated parameters for all epochs are listed in Table \ref{tab:en-pol}. Note that the estimated PD for some of the energy bands (see Fig. \ref{fig:ixpe_pol}b) remains very close to or below the $\rm MDP_{99}\%$ level due to poor statistics obtained in the respective energy bands, which requires careful interpretation of the obtained results. Furthermore, in Fig. \ref{fig:stokes}b, we show the variation of normalized Stokes parameters (Q/I and U/I) obtained for epoch IX1 over different energy bands with $1\sigma$ contours. As before, a measurement of the significance associated with the variation in Stokes parameter space results in a marginally improved statistical interpretation of the energy variation of polarimetric parameters, significant at $97.6\%$ (IX1), $95.4\%$ (IX2) and $84.4\%$ (IX3) confidence levels in different epochs.

\subsubsection{Model-dependent Spectro-polarimetric Results}

We simultaneously fit the I, Q and U Stokes spectra from all DUs ($2-8$ keV) within \texttt{XPSEC}, considering a standard model combination \texttt{const$\ast$Tbabs$\ast$polconst$\ast$(gaussian+bbodyrad+cutoffpl)}, for spectro-polarimetric modeling. Here, \texttt{polconst} represents constant polarization model with PD and PA as the model parameters. We obtain the best fit with $\chi_{\rm red}^{2}=1.2$ resulting in PD $=1.9 \pm 0.2\%$ and PA $=10.3^{\circ} \pm 2.9^{\circ}$ in epoch IX1. Further, the seed photon temperature ($kT_{\rm bb}$) and photon index ($\Gamma$) are obtained as $2.03_{-0.04}^{+0.05}$ keV and $1.64\pm0.03$, respectively, during epoch IX1.

Next, we replace \texttt{polconst} by the energy dependent polarization model \texttt{polpow} that fits Q and U Stokes spectra considering PD($E$)$= A_{\rm norm}\times E^{-A_{\rm index}}$ and PA($E$)$= \psi_{\rm norm}\times E^{-\psi_{\rm index}}$. We find that the model fitted $\psi_{\rm index}$ remains consistent with zero within $1\sigma$. Hence, we freeze it to zero while modelling. This yields the best fitted polarization parameters of epoch IX1 having $A_{\rm index}=-0.98\pm0.3$, $A_{\rm norm}=0.0051_{-0.0019}^{+0.0028}$ and $\psi_{\rm norm}=10.5^{\circ} \pm 2.7^{\circ}$ with $\chi_{\rm red}^{2}=1.2$. It is worth mentioning that $A_{\rm index}$ is consistent with zero at $3\sigma$ level, suggesting a marginal energy variation ($< 3\sigma$). This is consistent with the energy dependent results significant at $2.5\sigma$ (IX1), obtained from the model-independent approach. Moreover, to explore the covariance between $A_{\rm index}$ and $A_{\rm norm}$, we conduct an MCMC simulation in \texttt{XSPEC} using the Goodman-Weare algorithm \cite[]{Goodman-Weare2010} with a chain length of 200000. The obtained results are presented in the left panel of Fig. \ref{fig:spec-pol}, where strong covariance between $A_{\rm index}$ and $A_{\rm norm}$ is observed in their respective error regions, which limits the ability to independently constrain one parameter from the other. Next, integrating PD($E$) and PA($E$) with best-fitted model parameters over $2-8$ keV, we obtain PD and PA as $2.3 \pm 0.2\%$ and $10.5^{\circ} \pm 2.7^{\circ}$, respectively. Similar model combinations are also found to provide the best fit in epochs IX2 (PD $= 3.6 \pm 0.2 \%$, PA $= 0.8^{\circ} \pm 3.5^{\circ} $) and IX3 (PD $= 3.0 \pm 0.3\%$, PA$= 7.5^{\circ} \pm 4.9^{\circ} $). In particular, a fit with the \texttt{pollin} model (linearly varying polarization with energy) yields the parameters \texttt{A1} and \texttt{Aslope} being consistent with zero at the $1\sigma$ level, and hence fails to constrain the polarization parameters PD and PA.

Further, we adopt the model combination \texttt{const$\ast$Tbabs$\ast$(polconst$\ast$gaussian} \texttt{+polconst$\ast$bbodyrad\\+polconst$\ast$cutoffpl)} to estimate the polarization level of each model component. We observe that the best fit yields PD $=4.7\pm1.3\%$ and PA $=11.1^{\circ}\pm8.4^{\circ}$ with $\chi_{\rm red}^{2}=1.2$ associated with the \texttt{bbodyrad} component, whereas the \texttt{gaussian} and \texttt{cutoffpl} components remain unpolarized, with PD and PA unconstrained at $1\sigma$. All the model fitted and estimated parameters obtained from the spectro-polarimetric modelling are tabulated in Table \ref{tab:specpol}. The best-fitted I, Q and U Stokes spectra of {\it IXPE} in $2-8$ keV energy range and the corresponding residual variations are depicted in the right panel of Fig. \ref{fig:spec-pol}. In the figure, the dotted curves represent the model components (\texttt{bbodyrad}, \texttt{cutoffpl}, and \texttt{Gaussian}) obtained from the best fit of the I Stokes spectra. The dot-dashed curve corresponds to the best-fitted model of the Q and U Stokes spectra, while the dashed curve depicts the best-fitted effective model for the I Stokes spectra.

Furthermore, we attempt to deduce the correlation between PD (model-independent), luminosity, and the ratio of \texttt{bbodyrad} to \texttt{cutoffpl} fluxes ($F_{\rm BB}/F_{\rm PL}$) obtained with \texttt{cflux} from the modelling of {\it IXPE} spectra ($2-8$ keV). In Fig. \ref{fig:ixpe_pol}c, we present the variation of PD ($2-5$ keV and $5-8$ keV) with luminosity ($2-8$ keV). The color code denotes the flux ratio ($F_{\rm BB}/F_{\rm PL}$) computed using the \texttt{polpow} component in the respective energy bands. We observe that the PD increases marginally from epoch IX1 to epoch IX2 and subsequently drops in epoch IX3 as luminosity decreases $(1.68-0.35)\times 10^{37}$ erg $\rm s^{-1}$ (Fig. \ref{fig:ixpe_pol}c) and $F_{\rm BB}/F_{\rm PL}$ increases in the respective energy bands. Note that \texttt{bbodyrad} dominates ($1.5 \lesssim F_{\rm BB}/F_{\rm PL} \lesssim 3.4$) in $5-8$ keV band, whereas \texttt{cutoffpl} becomes prominent ($0.4 \lesssim F_{\rm BB}/F_{\rm PL} \lesssim 1$) in $2-5$ keV energy range. We observe that the PD reaches up to $\sim 3.0-4.8\%$ in $5-8$ keV for which \texttt{bbodyrad} flux exceeds twice that of \texttt{cutoffpl}. However, it is important to note that the change in PD over the epochs in different energy bands is marginally significant at $74\%$ ($2-5$ keV) and $73\%$ ($5-8$ keV) confidence levels, respectively (see Table \ref{tab:en-pol}). Therefore, the results of the correlation study presented here need to be interpreted with caution considering their statistical significance.

\subsection{Phase-resolved Polarimetric Properties} \label{s:phase_pol}

\begin{figure}
	\begin{center}
		\includegraphics[width=\columnwidth]{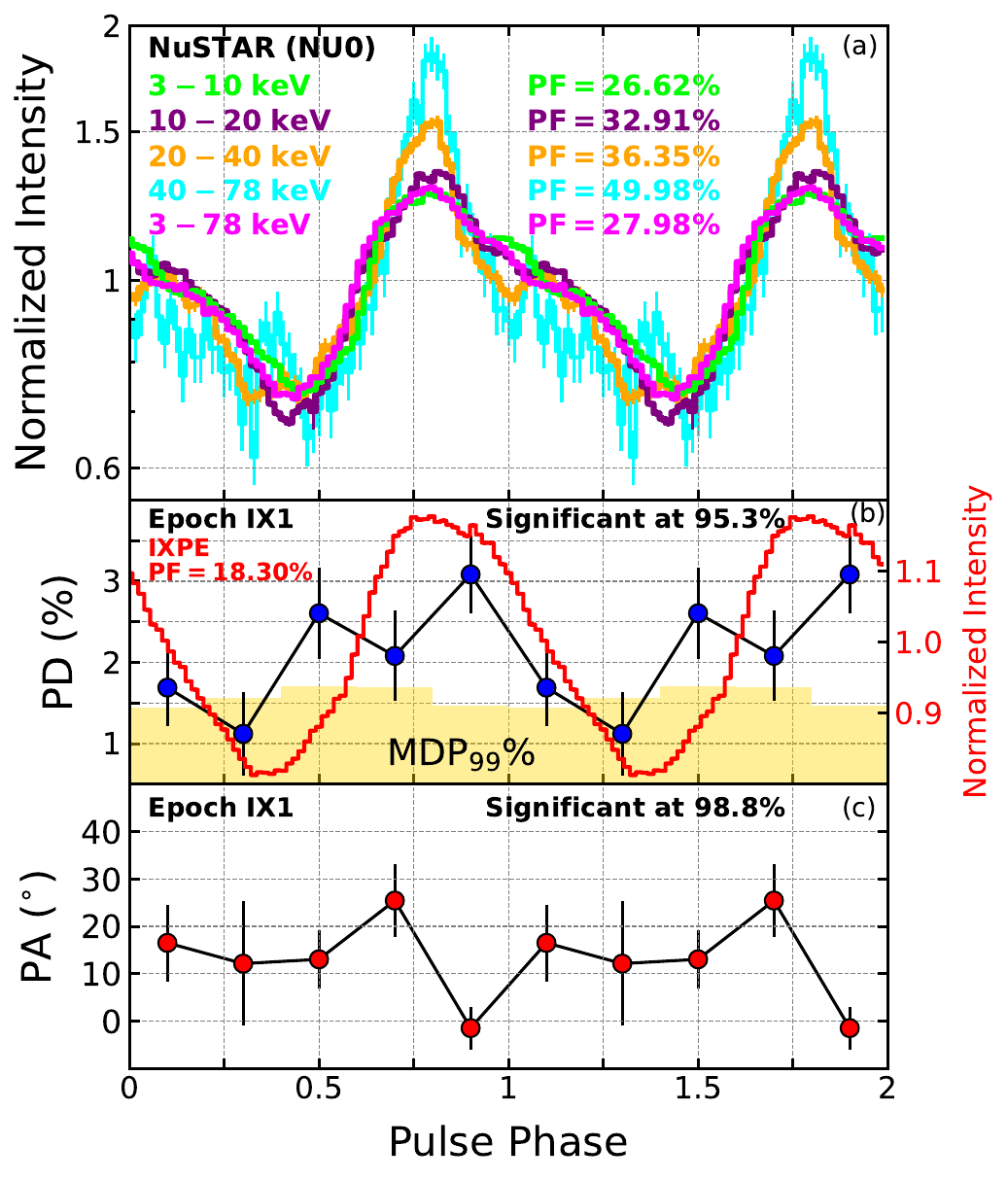}
	\end{center}
	\caption{Panel (a): Pulse profile of Swift J0243.6$+6124$ obtained from {\it NuSTAR} in different energy bands are depicted with different colors. Panel (b): Variation of PD and pulse profile with the pulse phase ($2-8$ keV) for epoch IX1. Histograms denote MDP$_{99}\%$. Panel(c): Variation of PA with pulse phase for epoch IX1. Note that the variations of PD and PA over different phase bins are marginal. See the text for details.
	}
	\label{fig:pulse_phase}
\end{figure}

We search for pulsation using epoch-folding \citep{leahy1983} in the $2-8$ keV \textit{IXPE} band, as well as in different energy bands ($3-10$ keV, $10-20$ keV, $20-40$ keV and $40-78$ keV) of {\it NuSTAR} data. The pulse profiles are normalized by dividing with the average intensity of the respective energy bands. Uncertainties in the pulse periods were determined by producing simulated light curves at each epoch, and computing the RMS variation in their determined periods (\citealt{leahy1987}; see also \citealt{chatterjee2021}).

The source exhibits strong pulsations in all \textit{NuSTAR} energy bands as well as in the entire energy band with P = 9.79909(4) s. The pulse profiles and pulse fractions (PFs) are found to be strongly energy dependent as PF increases from $\sim 27\%$ to $\sim 50\%$ with the increase in energy \cite[see][]{Beri-etal2021}. The pulse profiles are dominated by emission from a single peak at lower energies and changes to a double-peaked profile at higher energies ({}Chatterjee et al. 2024, in preparation). Pulsations are also detected in all the \textit{IXPE} epochs (see Table~\ref{tab:en-pol}) in $2-8$ keV with pulse profiles of single-peaked (IX1, $\rm  PF \sim 18\%$), double-peaked (IX2, fundamental $\rm  PF \sim 15\%$) and triple-peaked (IX3, fundamental $\rm  PF \sim 10\%$) nature, respectively, with decreasing luminosity. The pulse profiles obtained with {\it NuSTAR} and {\it IXPE} are shown in panels (a) and (b) of Fig. \ref{fig:pulse_phase}, respectively.

To investigate the phase-resolved polarization properties of Swift~J0243.6$+6124$, we divide the epoch IX1 observation into 5 equal phase-bins of width $0.2$ each and compute PD and PA in each bin, depicted in panels (b) and (c) of Fig. \ref{fig:pulse_phase}, respectively. We observe a moderate variation of PD with phase in the range $\sim 1.7-3.1\%$, exhibiting a possible correlation with intensity (see Fig. \ref{fig:pulse_phase}b). Similarly, PA also shows marginal variation within $12.2^{\circ}-25.5^{\circ}$ in the phase bins. In particular, we find that the variations of both PD and PA over different phase bins for IX1 are significant at $95.3\%$ and $98.8\%$ confidence levels against the phase-averaged values of $2-8$ keV energy range (see Table \ref{tab:en-pol}). However, we observe that the PD obtained in some phase bins remains close to or below the $\rm MDP_{99}\%$ level, mainly due to limited statistics across the phase bins\footnote{More than $5$ phase bins are avoided as the PD remains below the $\rm MDP_{99}\%$ level. See Appendix for details.}. In Fig. \ref{fig:stokes}c, we illustrate the variation of normalized Stokes parameters in multiple phase bins for epoch IX1. Following the approach mentioned in Section 3.2.1, we estimate the significance of the variation of the normalized Stokes parameters (Q/I and U/I) over five phase bins as $98.3\%$ for epoch IX1. Indeed, the variation of the polarimetric properties of the source remains more significant ($> 2 \sigma$) in Stokes parameter space as compared to the variation in PD ($< 2 \sigma$), as explained earlier.

\subsection{Broad-band Spectral Distribution} \label{s:spectra}

\begin{figure}
	\begin{center}
		\includegraphics[width=\columnwidth]{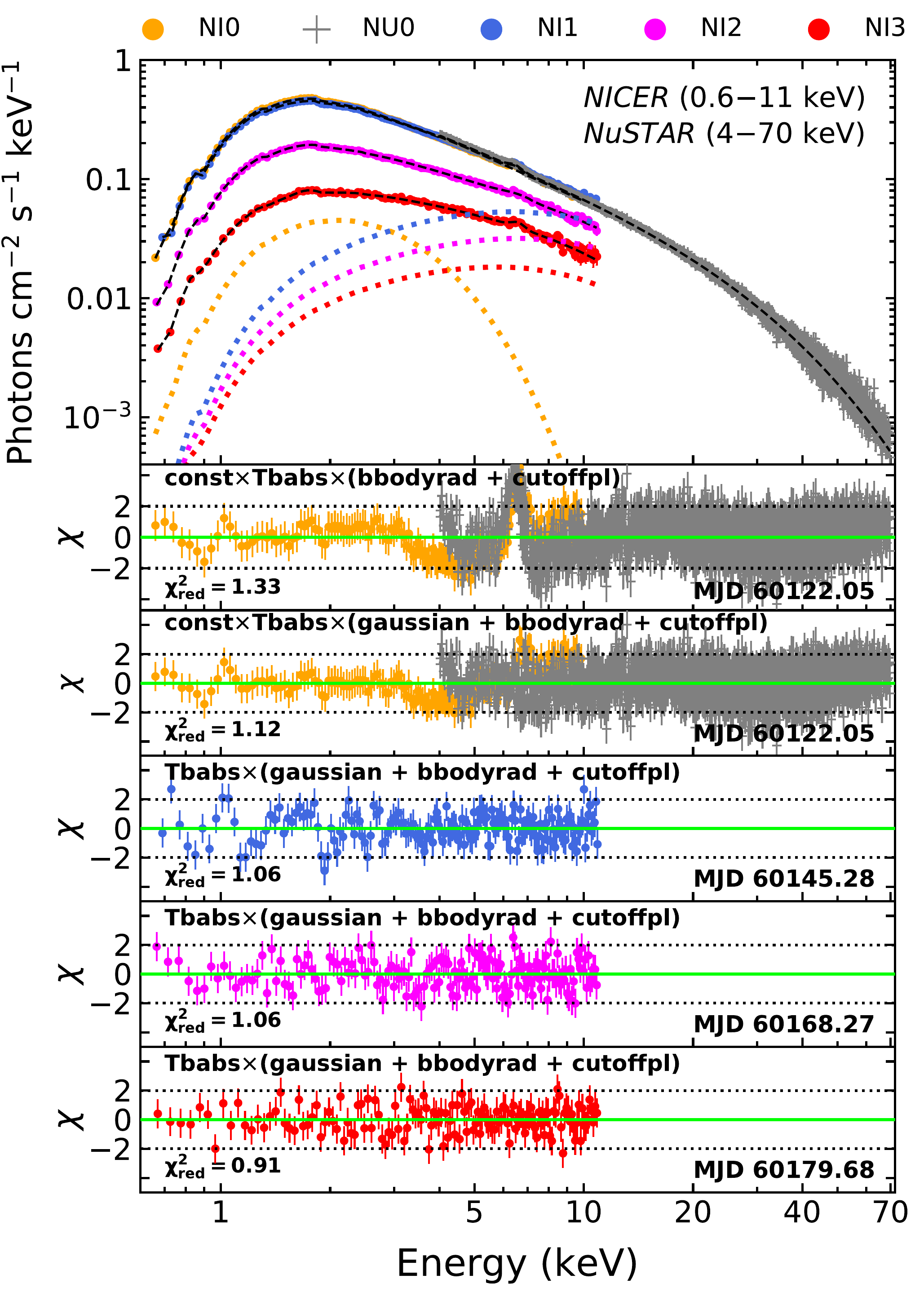}
	\end{center}
	\caption{Best fitted broad-band energy spectra of Swift J0243.6$+$6124 in $0.6-70$ keV energy band from quasi-simultaneous {\it NICER} and {\it NuSTAR} observations (NI0, NU0) and {\it NICER} spectra ($0.6-11$ keV) during epoch NI1, NI2 and NI3, respectively. See the text for details.}
		
	\label{fig:spec}
\end{figure}

We examine broad-band ($0.6-70$ keV) energy spectral energy distribution of Swift J0243.6$+6124$ using quasi-simultaneous {\it NICER} (NI0) and {\it NuSTAR} (NU0) observations. In general, accretion-powered pulsars exhibit emission over a wide range of energies, including soft X-ray peaks and hard tails at higher energies. We adopt a model combination \texttt{const$\ast$Tbabs$\ast$(gaussian+bbodyrad+cutoffpl)} comprising of a \texttt{bbodyrad} component for thermal emission along with a \texttt{cutoffpl}, accounted for higher energy cut-off, to model the broad-band continuum. The \texttt{gaussian} is used to model the strong iron line emission observed at $\sim 6.4$ keV and \texttt{Tbabs} \citep{Wilms-etal2000} takes care of the interstellar absorption. We find the best fit with the above model combination as $\chi^{2}_{\rm red} (\chi^{2}/d.o.f)$ $=1.13$ $(3136/2765)$. Further, we model the {\it NICER} spectra during epochs NI1, NI2 and NI3 in $0.6-11$ keV energy range. Note that a similar model combination provides an acceptable fit for all {\it NICER} spectra with $0.91 \le \chi^{2}_{\rm red} \le 1.06$.

The best fitted broad-band spectrum (NI0$+$NU0) results in seed photon temperature $kT_{\rm bb}=0.86\pm0.03$ keV with $norm_{\rm bbody}=135_{-20}^{+24}$. We find the photon index ($\Gamma$) of \texttt{cutoffpl} as $0.98\pm0.01$ with a high energy cut-off ($E_{\rm c}$) at $20.16\pm0.24$ keV. The hydrogen column density ($n_{\rm H}$) is found to be $(1.01\pm0.01) \times 10^{22}$ cm$^{-2}$. We detect iron line emission at $6.45\pm0.02$ keV of line width $0.24\pm0.03$ keV in NI0$+$NU0. In addition, $kT_{\rm bb}=3.50_{-0.34}^{+0.61}-3.92_{-0.15}^{+0.16}$ keV, $norm_{\rm bbody}=2.26_{-0.97}^{+1.07}-5.25_{-0.59}^{+0.64}$ and $\Gamma = 0.57_{-0.10}^{+0.11}-1.08_{-0.02}^{+0.02}$ are obtained during NI1-NI3 observations. Note that $E_{\rm c}$ remains unconstrained in the {\it NICER} spectra of NI1-NI3, and hence we freeze it at $6$ keV.

Using \texttt{cflux}, we obtain the bolometric flux ($0.1-100$ keV) as $3.57\times10^{-8}$ erg $\rm cm^{-2}$ $\rm s^{-1}$ in NI0$+$ NU0, which decreases sharply in the range $(2.75-0.72)\times 10^{-8}$ erg $\rm cm^{-2}$ $\rm s^{-1}$ during NI1 to NI3. Considering a distance of $6.8$ kpc to the source \citep{Bailer-Jones-etal2018}, we find the bolometric luminosity of the source as $1.98 \times 10^{38}$ erg $\rm s^{-1}$ for NI0$+$NU0, which exceeds the Eddington limit of XRPs, confirming its ultraluminous nature \cite[see also][]{Wilson-etal2018,Chhotaray-etal2024}. In Fig. \ref{fig:spec}, we show the best-fitted broad-band ($0.6-70$ keV) energy spectrum of combined NI0$+$NU0 observations along with {\it NICER} spectra ($0.6-11$ keV) during NI1-NI3 epochs. The dotted lines of various colors depict the best-fitted \texttt{bbodyrad} component corresponding to the respective spectra. It is worth noting that despite NI0 and NI1 observations being roughly a month apart, the spectral shapes appear to remain quite similar (see Fig. \ref{fig:swift_BAT} and Fig. \ref{fig:spec}).

\section{Discussion} \label{s:dis}

In this Letter, we report the results of spectro-polarimetric studies of the Galactic ULXP Swift J0243.6$+6124$ using first ever {\it IXPE} observations ($2-8$ keV) during the 2023 outburst. We carry out epoch-dependent spectro-polarimetric study with {\it IXPE} over three epochs (IX1, IX2 and IX3) with an integrated exposure of $\sim 375$ ks. 

Indeed, the main findings obtained in this work is the detection of significant polarized emission of PD $\sim 2.0\pm 0.2\%$ ($> 8\sigma$) and PA $\sim 10.8^{\circ} \pm 3.3^{\circ}$ in $2-8$ keV energy range during epoch IX1. This finding confirms Swift J0243.6$+6124$ as the first ULXP that exhibits the signature of polarized emissions. We find that the PD increases up to $\sim 3.1\%$ in epoch IX2 and eventually decreases to $\sim 2.4\%$ during epoch IX3. However, moderate variation of PD with energy reaching up to $\sim 3.3\%$ (IX1), $\sim 4.5\%$ (IX2) and $\sim 3.6\%$ (IX3) at $\sim 7$ keV (see Fig. \ref{fig:ixpe_pol}b and Table \ref{tab:en-pol}) is observed. Notably, no significant variation of PA is seen in different energy ranges for all the epochs. Needless to mention that the observed PD appears much lower compared to the theoretical model predictions (up to $\sim 80\%$) for XRPs \citep{Meszaros-etal1988,Caiazzo-etal2021b, Caiazzo-etal2021a}. Owing to that, the polarization results of Swift J0243.6$+6124$ are in agreement with the reported `low' PDs in other XRPs, namely Cen X$-1$ ($\sim 5.8\%$ \citealt{Tsygankov-etal2022}), 4U 1626$-67$ ($< 4\%$, \citealt{Marshall-etal2022}) and Vela X$-1$ ($\sim 2.3\%$, \citealt{Forsblom-etal2023}). 

We also find that the PD increases with the decrease in luminosity between epoch IX1 and IX2 (Fig. \ref{fig:ixpe_pol}c) and subsequently decreases at epoch IX3. Intriguingly, we observe an overall increase of PD as $3-4.8\%$ in the presence of a dominant \texttt{bbodyrad} emission ($1.5 \lesssim F_{\rm BB}/F_{\rm PL} \lesssim 3.4$) at higher energies ($\gtrsim 5$ keV) of the {\it IXPE} spectra. Nevertheless, PD remains within $1.65-2.42\%$ in \texttt{cutoffpl} dominated spectral domain ($0.4 \lesssim F_{\rm BB}/F_{\rm PL} \lesssim 1$) below $\sim 5$ keV (see Fig. \ref{fig:ixpe_pol}c). In addition, the spectra of quasi-simultaneous {\it NICER} epochs (NI1, NI2 and NI3) also show the dominance of the \texttt{bbodyrad} component beyond $\sim 5$ keV (see Fig. \ref{fig:spec}). Based on these findings, we infer that the \texttt{bbodyrad} emission contributes to the observed high polarization degree up to $\sim 7$ keV and the \texttt{cutoffpl} component (dominated in $\sim 2-5$ keV) results in a depolarization in $2-8$ keV energy band with PD $\sim 2-3.1\%$.

We detect strong pulsations in Swift J0243.6$+6124$ with a spin period $\sim 9.79$ s \cite[see][]{Serim-etal2023,Chhotaray-etal2024} during epoch IX1 in $2-8$ keV energy range having a pulse fraction (PF) of $\sim 18\%$ (Fig. \ref{fig:pulse_phase}b).  {\it NuSTAR} also detects $\sim 9.79$ s pulsation period ($3-78$ keV) with PF $\sim 28\%$. The Energy dependent pulsation study with {\it NuSTAR} further yields a monotonic increase of PF from $27\%$ ($3-10$ keV) to $50\%$ ($40-78$ keV). The phase-resolved polarimetric study reveals the variation of PD and PA within $\sim 1.7-3.1\%$ and $\sim 12.2^{\circ}-25.5^{\circ}$ with pulse phase, respectively. Interestingly, we observe a marginal correlation between PD and the pulse intensity with a minimal variation of PA (see Fig. \ref{fig:pulse_phase}c). However, an anti-correlation between PD and source intensity is reported for several XRPs including Cen X$-3$ \citep{Tsygankov-etal2022} and GRO J1008$-57$ \citep{Tsygankov-etal2023}. It is worth mentioning that, unlike the case of Swift J0243.6$+6124$, several pulsars exhibit relatively high PD ($\gtrsim 10\%$) along with significant PA variation as reported from phase dependent polarimetric studies \citep{Tsygankov-etal2022, Forsblom-etal2023, Suleimanov-etal2023, Tsygankov-etal2023}. Moreover, the wide variation of PA over phase bins in many of such pulsars is satisfactorily described using the rotating vector model \citep{Radhakrishnan-etal1969, Poutanen-etal2020}.

Meanwhile, various physical mechanisms are proposed to explain the relatively `low' PDs observed in most of the XRPs. These include (a) reflected fraction of hot spot emission from the NS surface, (b) accretion curtain and (c) accretion disc, all of which have the potential to produce substantial polarization \cite[]{Tsygankov-etal2022} depending on the magnetic field strength of the NS atmosphere \citep{Poutanen-etal1996}. In addition, the emissions reflected from the optical companion can alter the polarization state depending on the pulsar's beam pattern \cite[]{Tsygankov-etal2022}. However, this effect becomes significant only above $\sim 10$ keV, making its contribution negligible in the {\it IXPE} band. The pronounced variation of the polarization angle with pulse phases leads to seemingly `low' averaged PDs, despite significant polarization being evident in phase dependent estimates \citep{Suleimanov-etal2023}.

Alternatively, it is proposed that `vacuum resonance' at the transition point between the overheated upper NS surface and relatively cooler underlying layer could play a viable role \citep{Doroshenko-etal2022}. In particular, ordinary and extraordinary modes of polarization start to convert into each other while passing through the vacuum resonance, where the polarization contribution from plasma and vacuum birefringence becomes equal \citep{Lai-etal2002}. In other words, these two contributions acting against each other result in the depolarization of radiation. Note that this effect may be enhanced for the XRPs at critical luminosity with variations in the emission regions \cite[]{Doroshenko-etal2022,Tsygankov-etal2022}. Swift J0243.6$+6124$ having $\sim 10^{13}$ G surface magnetic field  \citep{Kong-etal2022} exhibits critical luminosity $\gtrsim 10^{37}$ erg $\rm s^{-1}$ \citep{Mushtukov-etal2015} which is close to the estimated luminosities during {\it IXPE} campaign. Hence, we speculate that the observed `low' PD in Swift J0243.6$+6124$ possibly resulted because of the vacuum resonance similar to the predictions for Her X$-1$ \citep{Doroshenko-etal2022} and Cen X$-1$ \citep{Tsygankov-etal2022}. However, the possibility to explain the observed `low' PD using alternative physical mechanisms cannot be entirely ruled out.

The broad-band energy spectrum ($0.6-70$ keV) with {\it NICER} and {\it NuSTAR} during NI0$+$NU0, described by \texttt{bbodyrad} and \texttt{cutoffpl} components indicates bolometric luminosity ($0.1-100$ keV) as $1.98\times 10^{38}$ erg $\rm s^{-1}$. The bolometric luminosity is seen to decrease as $1.52-0.40\times 10^{38}$ erg $\rm s^{-1}$ in the {\it NICER} epochs (NI1, NI2 and NI3) during the decay phase of the outburst. A substantially higher luminosity of approximately $\sim 40 L_{\rm Edd}$ was observed during the source's 2017 outburst, confirming its ultra-luminous nature \citep{Tsygankov-etal2018}.

In conclusion, we report the first detection of phase-averaged as well as phase-resolved polarization in the Galactic X-ray pulsar Swift J0243.6$+6124$ during the decaying phase of the 2023 outburst. 

\section{Acknowledgments}

Authors thank the anonymous reviewer for constructive comments and useful suggestions that helped to improve the quality of the manuscript. SD thanks Science and Engineering Research Board (SERB) of India for support under grant MTR/2020/000331. RC, KJ and AN thank GH, SAG; DD, PDMSA, and Director, URSC for encouragement and continuous support to carry out this research. This publication uses data from the {\it IXPE}, {\it NICER} and {\it NuSTAR} missions. This research has made use of the {\it MAXI} data provided by {\it RIKEN}, {\it JAXA} and the {\it MAXI} team \citep{Matsuoka-etal2009}. The {\it Swift/BAT} transient monitor results provided by the {\it Swift/BAT} team are also used \citep{Krimm-etal2013}. We thank each instrument team for processing the data and providing necessary software tools for the analysis.

\section{Data Availability}

Data used for this publication are currently available at the HEASARC browse website (\url{https://heasarc.gsfc.nasa.gov/db-perl/W3Browse/w3browse.pl}).

	
	\begin{table*}
		\caption{Details of quasi-simultaneous {\it IXPE}, {\it NICER} and {\it NuSTAR} observations of Swift J0243.6$+6124$.}
		
		\renewcommand{\arraystretch}{1.2}
		\resizebox{0.8\textwidth}{!}{%
			\begin{tabular}{l l c l c c c}
				\hline
				
				Epoch & Mission & Date & ObsID & MJD$_{\rm start}$ & MJD$_{\rm stop}$ & Exposure (ks) \\
				
				\hline
				
				NI0 & NICER & 27-06-2023 & 6050390237 & 60122.05 & 60122.96 & 4 \\
				
				NU0 & NuSTAR & 27-06-2023 & 90901321002 & 60122.53 & 60123.08 & 12 \\
				
				IX1 & IXPE & 20-07-2023 & 02250799 & 60145.63 & 60148.65 & 167 \\
				
				NI1 & NICER & 20-07-2023 & 6050390252 & 60145.28 & 60145.93 & 1.6 \\
				
				& NICER & 21-07-2023 & 6050390253 & 60146.51 & 60146.77 & 0.8\\
				    
				& NICER & 22-07-2023 & 6050390254 & 60147.02 & 60147.80 & 1.9 \\
				
				IX2 & IXPE & 09-08-2023 & 02250799 & 60165.99 & 60167.37 & 77 \\
				
				NI2 & NICER & 12-08-2023 & 6050390255 & 60168.27 & 	60168.98 & 1.7 \\
				
				IX3 & IXPE & 23-08-2023  & 02250799 & 60179.42 & 60181.81 & 131 \\
				
				NI3 & NICER & 23-08-2023 & 6050390265 & 60179.68 & 	60179.69 & 0.7 \\

				\hline
			\end{tabular}%
		}
		\label{table:obs_log}
	\end{table*}


\begin{table*}
	\centering
        
        \caption{Results from model-independent polarimetric analyses in different energy bands for three epochs (IX1, IX2 and IX3). Here, PD, PA, Q/I, U/I, $\rm MDP_{99}$ and SIGNIF denote polarization degree, angle of polarization, normalized Q-Stokes parameter, normalized U-Stokes parameter, minimum detectable polarization at $99\%$ confidence and detection significance, respectively.}
        
		\renewcommand{\arraystretch}{1.2}
		\resizebox{1.0\textwidth}{!}{%
  	\begin{tabular}{l c l c c c c c c c c c c}
		\hline
		
		Epoch & Pulse Period & Parameters &  $2-3.5$  & $3.5-5$ & $5-6.5$  & $6.5-8$  & $2-8$ & $2-5$ & $5-8$ \\
		(Exposure) & (s) &  & (keV)  & (keV) & (keV) & (keV) & (keV) & (keV) & (keV) \\
		\hline

		IX1 & $9.79301(1)$ &  PD(\%)  & $1.4 \pm 0.3$ & $2.0 \pm 0.3$ & $3.3 \pm 0.5$ &  $2.5 \pm 1.0$ &  $2.0 \pm 0.2$  & $1.7 \pm 0.2$ & $3.0 \pm 0.5$ \\
		
		($\sim 167$ ks) & &  PA ($^\circ$)  & $9.0 \pm 5.1$ & $14.6 \pm 4.6$ & $7.1 \pm 4.6$ &  $15.6 \pm 12.0$ &  $10.8 \pm 3.3$  & $11.5 \pm 3.6$ & $9.7 \pm 5.1$ \\
		
		&  & Q$/$I (\%)  & $1.4 \pm 0.3$ & $1.8 \pm 0.3$ & $3.2 \pm 0.5$ &  $2.1 \pm 1.0$ &  $1.9 \pm 0.2$  & $1.5 \pm 0.2$ & $2.8 \pm 0.5$ \\
		
		&  & U$/$I (\%)  & $0.4 \pm 0.3$ & $1.0 \pm 0.3$ & $0.8 \pm 0.5$ &  $1.3 \pm 1.0$ &  $0.7 \pm 0.2$  & $0.6 \pm 0.2$ & $1.0 \pm 0.5$ \\
		
		&  & $\rm MDP_{99}$ (\%)  & $0.8$ & $1.0$ & $1.6$ &  $3.1$ &  $0.7$  & $0.6$ & $1.6$ \\
		
		&  & SIGNIF ($\sigma$)  & $5.1$ &  $5.7$ &  $5.8$ &  $1.6$ &   $8.7$  &  $7.6$ &  5.2  \\
		
		\hline
		
		IX2 & $9.79204(2)$ & PD(\%)  & $2.2 \pm 0.5$ & $2.9 \pm 0.6$ & $4.5 \pm 1.0$ &  $5.2 \pm 1.9$ &  $3.1 \pm 0.5$ & $2.4 \pm 0.4$ & $4.8 \pm 1.0$ \\
		
		($\sim 77$ ks) &  & PA ($^\circ$)  & $-0.8 \pm 6.8$ & $-0.6 \pm 6.4$ & $-1.2 \pm 6.2$ &  $7.3 \pm 10.1$ &  $0.7 \pm 4.2$  & $-0.7 \pm 4.8$ & $2.5 \pm 5.8$ \\
		
		&  & Q$/$I (\%)  & $2.2 \pm 0.5$ & $2.9 \pm 0.6$ & $4.5 \pm 1.0$ &  $5.1 \pm 1.9$ &  $3.1 \pm 0.5$  & $2.4 \pm 0.4$ & $4.7 \pm 1.0$ \\
		
		&  & U$/$I (\%)  & $-0.1 \pm 0.5$ & $-0.1 \pm 0.6$ & $-0.2 \pm 1.0$ &  $1.3 \pm 1.9$ &  $0.1 \pm 0.5$  & $-0.1 \pm 0.4$ & $0.4 \pm 1.0$ \\
		
		&  & $\rm MDP_{99}$ (\%)  & $1.6$ & $1.9$ & $3.0$ &  $5.6$ &  $1.4$  & $1.2$ & $2.9$ \\
		
		&  & SIGNIF ($\sigma$)  & $3.6$ &  $3.9$ &  $4.1$ &  $2.1$ &  $6.5$  & $5.50$ & $4.43$ \\
		
		\hline
		
		IX3 & $9.79371(3)$ & PD(\%)  & $1.4 \pm 0.6$ & $2.8 \pm 0.7$ & $3.6 \pm 1.1$ &  $3.5 \pm 2.0$ &  $2.4 \pm 0.5$  & $2.0 \pm 0.5$ & $3.4 \pm 1.1$ \\
		
		($\sim 131$ ks) &  & PA ($^\circ$)  & $-2.8 \pm 12.6$ & $5.1 \pm 7.5$ & $10.0 \pm 8.7$ &  $26.1 \pm 16.5$ &  $8.6 \pm 6.4$  & $1.8 \pm 7.0$ & $16.2 \pm 8.9$ \\
		
		&  & Q$/$I (\%)  & $1.4 \pm 0.6$ & $2.7 \pm 0.7$ & $3.4 \pm 1.1$ &  $2.2 \pm 2.0$ &  $2.3 \pm 0.5$  & $2.0 \pm 0.5$ & $2.9 \pm 1.1$ \\
		
		&  & U$/$I (\%)  & $-0.1 \pm 0.6$ & $0.5 \pm 0.7$ & $1.2 \pm 1.1$ &  $2.8 \pm 2.0$ &  $0.7 \pm 0.5$  & $0.1 \pm 0.5$ & $1.8 \pm 1.1$ \\
		
		&  & $\rm MDP_{99}$ (\%)  & $1.9$ & $2.2$ & $3.3$ &  $6.2$ &  $1.6$  & $1.5$ & $3.2$ \\
		
		&  & SIGNIF ($\sigma$)  &  $1.4$ &  $3.2$ &  $2.6$ &  $0.8$ &  $3.9$  & $3.5$ & $2.5$ \\
		
		\hline

	\end{tabular}
		}
	
	\label{tab:en-pol}
\end{table*}


\begin{table*}
	\caption{Results from spectro-polarimetric analysis of the {\it IXPE} Stokes spectra for different epochs in $2-8$ keV energy range.}
	\resizebox{1.02\textwidth}{!}{%
		\begin{tabular}{l c c c c c c c c c c c c}
			\hline
			
			Component & Epoch & $A_{\rm norm}$  & $A_{\rm index}$ & $\psi_{\rm norm}$ & $kT_{\rm bb}$ & $\Gamma$ & $\chi^{2}_{\rm red}$ & $\rm PD$ & $\rm PA$ & $F_{\rm BB}/F_{\rm PL}$ & $L_{\rm IXPE}$ ($2-8$ keV) & $L_{\rm NICER}$ ($0.6-11$ keV) \\
			
			& & & &  ($^\circ$) & (keV) & & & ($\%$) & ($^\circ$) & ($2-8$ keV) & ($\times 10^{37}$ erg s$^{-1}$) & ($\times 10^{37}$ erg s$^{-1}$) \\ \hline
			
			\texttt{polconst} & IX1 & $-$ & $-$ & $-$ & $2.03_{-0.04}^{+0.05}$ & $1.64_{-0.03}^{+0.03}$ & $1.20$ & $1.86\pm0.19$ & $10.28 \pm 2.86$ &  $-$ & $-$ & $-$ \\ 
			
			& IX2 & $-$ & $-$ & $-$ & $2.12_{-0.08}^{+0.09}$ & $1.66_{-0.07}^{+0.07}$ & $1.15$ & $2.81\pm0.36$ & $0.70 \pm 3.68$ &  $-$ & $-$ & $-$ \\
			
			& IX3 & $-$ & $-$ & $-$ & $2.03_{-0.07}^{+0.08}$ & $1.46_{-0.13}^{+0.13}$ & $1.02$ & $2.44\pm0.42$ & $5.93 \pm 4.97$ &  $-$ & $-$ & $-$ \\
			
			\hline
			
			\texttt{polpow}$^{\dagger}$ & IX1 & $0.0051_{-0.0019}^{+0.0028}$ & $-0.98_{-0.30}^{+0.30}$ & $10.46 \pm 2.74$ & $2.03_{-0.04}^{+0.05}$ & $1.64_{-0.03}^{+0.03}$ & $1.20$ & $2.30\pm0.23$ & $10.46 \pm 2.74$ &  $0.71$ & $1.68 \pm 0.02$ & $8.58 \pm 0.02$ \\ 
			
			& IX2 & $0.0075_{-0.0036}^{+0.0062}$ & $-0.98_{-0.42}^{+0.41}$ & $0.77 \pm 3.53$ & $2.12_{-0.08}^{+0.09}$ & $1.66_{-0.07}^{+0.07}$ & $1.14$ & $3.63 \pm 0.17$ & $0.77 \pm 3.53$ & $1.10$ & $0.46 \pm 0.01$ & $4.37 \pm 0.02$ \\ 
			
			& IX3 & $0.0084_{-0.0049}^{+0.0056}$ & $-0.79_{-0.56}^{+0.54}$ & $7.46 \pm 4.93$ & $2.03_{-0.07}^{+0.08}$ & $1.46_{-0.13}^{+0.13}$ & $1.02$ & $2.96 \pm 0.31$ & $7.46 \pm 4.93$ & $1.73$ & $0.35\pm0.01$ & $2.21 \pm 0.01$ \\ 
			
			\hline
			
		\end{tabular}
	}
	
	\begin{list}{}{}
			\item $^{\dagger}$$\psi_{\rm index}$ of \texttt{polpow} is consistent with zero at $1\sigma$. Hence, it is frozen to zero while obtaining best fit parameters.
	\end{list}
	
	\label{tab:specpol}
\end{table*}


\appendix

\section{Pulse-phase dependent analysis}
\restartappendixnumbering

\begin{figure}
	\begin{center}
		\includegraphics[width=0.4\textwidth]{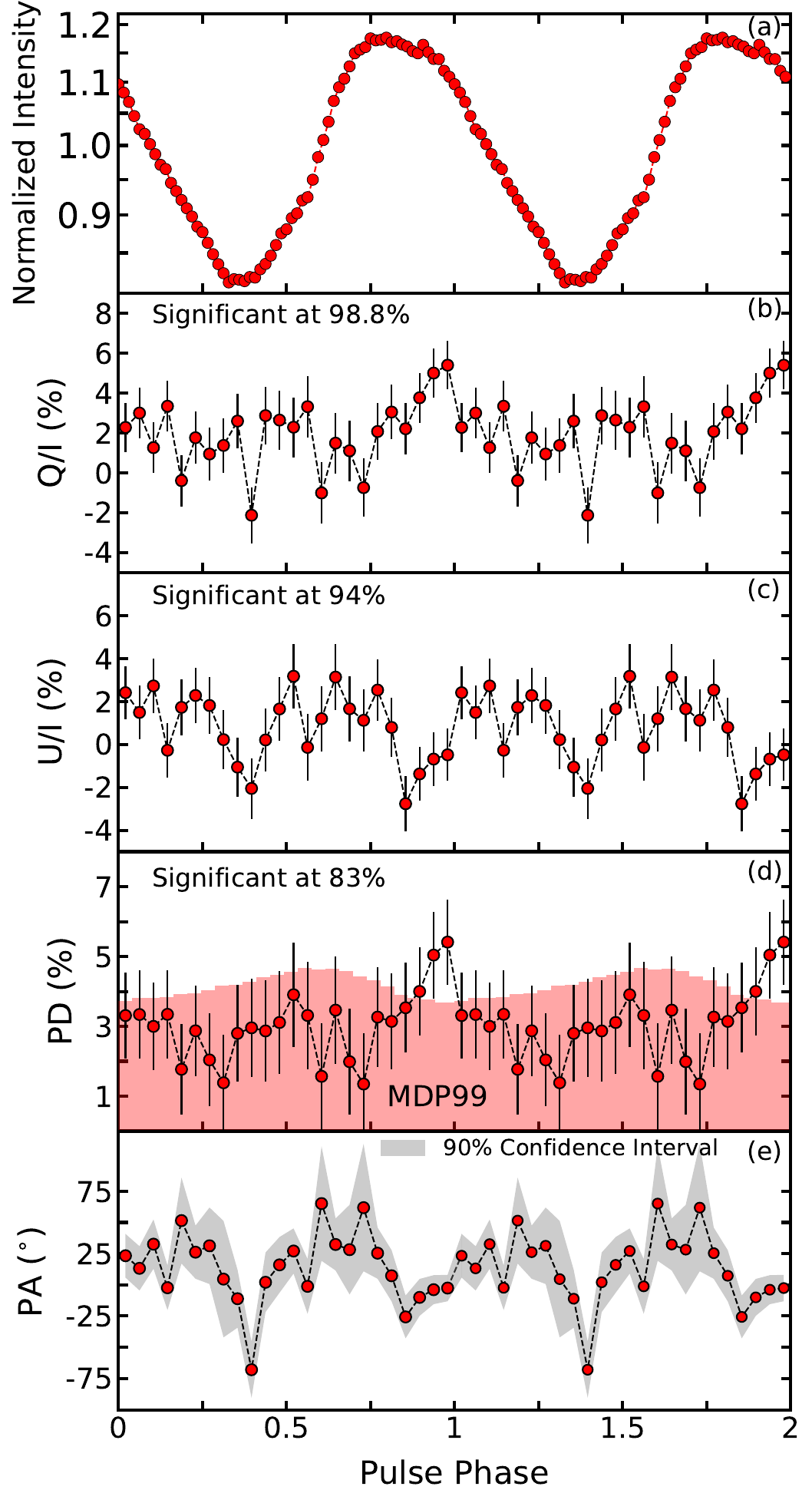}
	\end{center}
	\caption{Phase-resolved polarimetric analysis of Swift J$0243.6+6124$ during epoch IX1 in $3-8$ keV energy range over $24$ phase bins. In panels (a-e), variation of normalized intensity, normalized Stokes parameters (Q/I and U/I), PD and PA with pulse phase are shown. In panel (d), the histograms denote $\rm MDP_{99}\%$ level. In panel (e), gray shade represents $90\%$ confidence intervals. See the text for details.}
	\label{fig:A1}
\end{figure}

To investigate the finer phase variations of the polarimetric properties of Swift J$0243.6+6124$, we repeat our phase-resolved polarimetric analysis considering a higher number of phase bins for epoch IX1. Barycentric correction as well as correction for the binary orbit of the source using its orbital parameters\footnote{https://gammaray.nsstc.nasa.gov/gbm/science/pulsars/lightcurves/swiftj0243.html} \cite[]{Malacaria-etal2020} are applied to the events. We divide the data of $3-8$ keV energy band into $24$ phase bins following the recent analysis by \cite{Poutanen-etal2024}, which is related to the present paper. The obtained results are depicted in Fig. \ref{fig:A1}, where we present the variation of normalized intensity, normalized Stokes parameters (Q/I and U/I), PD and PA in panels (a)-(e), respectively. We observe moderate variation in Q/I and U/I, significant at $98.8\%$ and $94\%$ confidence levels, respectively, over the phase bins. The PD is found to vary in the range of $\sim 1-5.5\%$, significant at $83\%$ confidence level, although remaining below the $\rm MDP_{99}\%$ level in most of the phase bins (see panel (d)). This could occur as a result of insufficient photon statistics within the respective phase bins. Furthermore, we observe a noticeable variation of PA as $-68^{\circ}\lesssim \rm PA \lesssim 65^{\circ}$ over different phase bins of IX1 epoch. It is worth mentioning that as the PD measurements are below the $\rm MDP_{99}\%$ level, the corresponding PA variations deserve careful consideration.

\label{lastpage}

\end{document}